\documentclass[useAMS,usenatbib]{mn2e}
\usepackage[dvips]{epsfig}
\title[]{The Interaction Of Multiple Convection Zones In A-type Stars}
\author[L.\ J.\ Silvers \& M.\ R.\ E.\ Proctor]{L.\ J.\ Silvers $^{1,2}$\thanks{E-mail:
ljs53@damtp.cam.ac.uk (LJS); mrep@damtp.cam.ac.uk (MREP)} and M.\ R.\ E.\ Proctor$^{1}$\footnotemark[1]\\
$^{1}$Department of Applied Mathematics and Theoretical
  Physics,University of Cambridge, Cambridge, CB3 OWA, United Kingdom\\
$^{2}$Laboratoire de Radioastronomie, D\'epartement de Physique,
Ecole
  Normale Sup\'erieure, Paris, 75231, France}
\begin{document}



\maketitle

\label{firstpage}

\begin{abstract}
A-type stars have a complex internal structure with the possibility of multiple
convection zones. If not sufficiently separated, such zones
will interact through the convectively stable regions that lie
between them. It is therefore of interest to ask whether the typical
conditions that exist within such stars are such that these
convection zones can ever be considered as disjoint.

In this paper we present results from numerical simulations that
help in understanding how increasing the distance between the
convectively unstable regions affects the interaction. We go on to
discuss the effect of varying the stiffness of the stable layer that
lies between the unstable regions. We show that in A-type stars the
convectively unstable regions are likely to interact through the
stable region that separates them. This has profound implications
for mixing and transport within these stars.
\end{abstract}


\section{Introduction}

The internal structure of main sequence stars varies greatly according to their spectral type (\cite{Schwarzschild}). For example, stars like the Sun typically have a radiative core with a superadiabatic, convecting outer region. On the other hand upper main sequence stars, including A-type stars, are frequently pictured to consist of a
radiative outer layer that surrounds a large convecting hydrogen burning core. Such variation in structure is important as it can give rise to
drastically different transport mixing rates of both passive and
dynamic quantities within these stars. In addition, the connection to
the stellar atmosphere, which lies above the photosphere, is
fundamentally different if the outermost part of the interior is
convectively stable or convectively unstable.

Interactions within A-type stars also lead to different chemical balances, which can give rise to not just one but
multiple convection zones for main-sequence
stars of this type. Observations
indicate that, near to the surface of these stars there is
at least one convection zone  (\cite{Landstreet}; \cite{STKLS}). Further
considerations of a theoretical nature predict at least one further
convection zone near the surface (\cite{TZLS}; \cite{Kupka2}). Such
zones are believed to be thin (if compared to the radius of the
star) yet important from the point of view of transportation  and mixing. Their properties have been
discussed for several decades (\cite{Siedentopf}; \cite{LTZ};
\cite{Kupka}). 
The outer convection zone in these stars, immediately below the
surface, is caused by the partial ionization of hydrogen and the
single ionization of helium. The lower convection zone is the result
of the second ionization of helium (\cite{LTZ}; \cite{Kupka}) and at
least one further convection zone has been postulated
(\cite{Kupka2}). There has been a long-standing debate about the nature or even existence of the inner convection zone(s); in particular whether it can be consistent with the gravitational settling of heavier elements (see for example the papers of \cite{Vauclair1974} and \cite{Richer1992}). We do not attempt to address these questions in the present study; rather, we focus on the basic problem; given that two layers can at least plausibly coexist, what factors influence their interaction, and to what extent do earlier simplified models give a correct picture of this interaction?

It is now well known that in stars, without the presence of rigid
boundaries, ascending and descending convectively driven motions
overshoot the layer that is convectively unstable. In the Sun (a
G-type star) such overshooting is observed at the solar photosphere
and is witnessed in the form of granulation. Further, such
overshooting occurs at the bottom of the solar
convection zone and is believed to play an important role in the
solar dynamo because it transports magnetic field into the solar tachocline (\cite{TBCT1}; \cite{TBC2});
this is an important part of the interface dynamo model originally proposed by Parker
(\cite{Parker}). Such overshooting behaviour will naturally occur at other similar interfaces. When there are multiple convection zones such overshooting leads to enhanced communication and transport between the unstable layers. It is of importance to understand the nature of this interaction.

In A-type stars in particular, with two convection zones that are in quite close proximity, fascinating dynamics and mixing may occur. Overshooting plumes from an
upper convection zone and a lower convection zone can interact in
the convectively stable region separating them. Furthermore if
conditions are right it is possible for plumes to overshoot
completely and pierce the other convection zone, which would lead to
transportation of `contaminants' directly from one convectively
unstable region into the other. Therefore, the two fundamental questions are: How far apart do these layers need to be
before they can be considered as disjoint; and how close must they be for direct penetration from one layer to the other to occur.

Earlier analytic work (\cite{TZLS}; \cite{LTZ}) on this problem adopted a mean-field approach, which gives a highly simplified view of the nonlinear interactions but allows the reduction of the problem to a relatively simple set of ODE's, and it serves as  a guide for the fully compressible
simulations that are the subject of this paper. Latour \textit{et
al.} concluded that the two convectively unstable layers need to be
separated by a distance of at least two pressure scale heights for
there to be no interaction between the layers. Such a condition can
be achieved in a number of ways via the variation of the different
parameters that naturally occur in the model. One way is to increase
the vertical extent of the domain and so increase the width of the
intermediate layer. Another way is to vary the conductivity of the
mid layer. In this paper we choose to explore the effects of both of
these changes given
that the separation of the two zones as well as their relative
conductivities can be different in different stars. For simplicity we choose
to focus on convective layers of fixed width but we note that one could alternatively 
fix the domain height and decrease the
width of the convection zones and so increase the width of the
convectively stable region; this was the approach adopted in a preliminary
investigation by Muthsam, Wolfgang, Friedrich $\&$
Liebich (\cite{MWFL}).

Muthsam \textit{et al.} examined  three cases in a three dimensional
model in a Cartesian geometry with a small aspect ratio. In this
simple study they showed that bringing the two convection zones
closer together, by shrinking the width of the convectively
unstable region, led to the convection layers merging as the
interaction between the layers increased. However, their preliminary
investigation warrants a more detailed study for a number of
reasons. First, while they did make a passing remark as to the
pressure scale height change across the box they did not comment on
how the pressure scale height changes across the mid-layer, which
Toomre \textit{et al.} indicated was the important factor. Further,
while they acknowledge the earlier work by Toomre \textit{et al.}
they did not relate their numerical calculation directly to that work.

This paper is organised as follows: In the next section we provide a
detailed discussion of our model, the numerical method used to
solve the equations and the parameters that we select. In section 3
we examine the effect of varying the mid-layer thickness and
stiffness of the convectively stable region before concluding in
section 4.

\section[]{Model}

We consider the evolution of a compressible fluid in a layer and consider a model that is in the spirit of earlier papers
on penetrative convection (\cite{TBCT1}; \cite{TBC2}); these in
turn represent a simple extension of studies of  convection in a single Cartesian
layer. The following scalings are used to express the equations in
dimensionless form  (\cite{MPW}): lengths are scaled with the layer
depth $d$, times with the isothermal sound travel time
$d/\sqrt{R_*T_0}$, density with its value at the top of the layer
$\rho_0$, temperature with its value at the top of the layer $T_0$.

The governing equations can then be expressed as:
\begin{equation}
\frac{\partial{\rho}}{\partial{t}}+\mathbf{\nabla}.
\rho\textbf{u}=0, \nonumber
\end{equation}
\begin{equation}
\rho\left(\frac{\partial \textbf{u}}{\partial
t}+\textbf{u}.\mathbf{\nabla}\textbf{u} \right) = -\mathbf{\nabla} P
+ \theta (m+1)\rho \hat{\textbf{z}}+\sigma\kappa\mathbf{\nabla} . \rho
\mathbf{\tau},  \nonumber
\end{equation}
\begin{equation}
\frac{\partial{T}}{\partial{t}}+ \textbf{u}. \mathbf{\nabla} T  = -
(\gamma -1) T \mathbf{\nabla}. \textbf{u}  +
\frac{\kappa(\gamma-1)\sigma \tau^{2}}{2}
+\frac{\gamma \kappa}{\rho} \nabla^{2} T,
\end{equation}
 where $z$ is taken downward, $\theta$ is the dimensionless temperature difference across
the layer, $R_*$ is the gas constant, $m $ is the polytropic index,
$\kappa=K/d \rho_{0} c_{P} \sqrt(R_{*} T_{0})$ is the dimensionless
thermal diffusivity, $\gamma$ is the ratio of specific heats,
$\mathbf{\tau}$ is the stress tensor given by:
\begin{equation}
\tau_{ij}=\frac{\partial{u_{i}}}{\partial{x_{j}}}+\frac{\partial{u_{j}}}{\partial{x_{i}}}-\frac{2}{3}\delta_{ij}\frac{\partial{u_{k}}}{\partial{x_{k}}},
\nonumber
\end{equation}
$P = \rho T$ and $\sigma$ is the Prandtl number.

In order to achieve the required basic state we allow the thermal profile to be non-linear and we take
\begin{eqnarray}
K & = & \frac{K_1}{2} \Big[1+\frac{K_2+K_3}{K_1}-\tanh
\Big(\frac{z-1}{\Delta}\Big) \nonumber\\ &+ &
\frac{K_3}{K_1}\tanh\Big(\frac{z-\mathcal{D}+1}{\Delta}\Big) \nonumber \\
&-&\frac{K_2}{K_1}\tanh\big(\frac{z-\mathcal{D}+1}{\Delta}\Big)\tanh\big(\frac{z-1}{\Delta}\Big)\Big]
\end{eqnarray}
\noindent where $\Delta$ is the characteristic size of the
transition region between each of the layers. In this work  the
characteristic sizes of the transition regions are taken to be the same for
simplicity. The static density and temperature profiles are found by
solving the equations of hydrostatic balance. To this
static state, throughout the domain, random perturbations are
introduced, with amplitudes which lie within the interval [-0.05,0.05].

The aspect ratio for the computational domain in this study is
4:4:$\mathcal{D}$, where $\mathcal{D}$ is the total depth of the
box, and the domain is assumed to be periodic in \textit{x} and
\textit{y}. The conditions on the upper and
lower boundaries are:

\begin{equation}
T= 1,\hspace{0.5cm} u_z=0, \hspace{0.5cm} \frac{\partial
u_x}{\partial z}= 0 \hspace{0.5cm} at \hspace{0.1cm} z=0. \nonumber
\end{equation}

\begin{equation}
\frac{\partial T}{\partial z} = \theta ,\hspace{0.5cm} u_z=0,
\hspace{0.5cm} \frac{\partial u_x}{\partial z}= 0. \hspace{0.5cm} at
\hspace{0.1cm} z=\mathcal{D}. \nonumber
\end{equation}

The governing equations above are solved using a parallel hybrid
finite-difference/pseudo-spectral code; the most comprehensive
description of the code can be found in (\cite{MPW}). Nonlinear
products are performed in configuration space, the transformation
from phase space being facilitated by fast fourier transforms.
Time-stepping is carried out in configuration space and an explicit
second-order Adams-Bashforth scheme is used with variable weights to
accommodate adaptive step-sizes.

The system we study has a large number of dimensionless parameters, making it impractical to conduct a complete survey. Thus a number are held fixed at values shown in Table
\ref{t1}. These parameters have been chosen so that we have time
dependent, highly supercritical convection occurring in both convection zones. Of course, stellar convection operates in a highly
turbulent regime. The computational cost of simulating fully turbulent convection at very high Reynolds number is presently prohibitive. Nonetheless our flows are fully time-dependent and possess sufficient spatial complexity for our purposes.

We use a subscript 1 on quantities relating to the upper convection
zone. Similarly, subscript 2 will be used to denote quantities for
the convectively stable layer and 3 to denote quantities in the
lower convective zone.

\begin{table}
\begin{center}
\begin{tabular}{|c|c|c|}
\hline
Param.   &   Description &  Value  \\
\hline
$\sigma$  &  Prandtl  Number  &  1.0 \\
$m_1=m_3$   &  Polytropic  Index in layers 1 and 3 &  1.0  \\
$\theta$  &  Thermal  Stratification  & 10\\
$\gamma$  &  Ratio  of  Specific  Heats & $5/3$ \\
$R_a$  &  Rayleigh number& $1.7 \times 10^5$   \\

\hline
\end{tabular}
\caption{The parameter definitions and values.} \label{t1}
\end{center}
\end{table}

The stiffness parameter, $S$, provides a useful measure of the
relative conductivities in this problem (for more detailed
discussion see, \cite{HTMZ}; \cite{TBCT1}). $S_2$ and $S_3$ are
related to the various polytropic indexes that appear in the problem
via:
\begin{equation}
S_2=\frac{m_2-m_{ad}}{m_{ad}-m_1}
\end{equation}
\begin{equation}
S_3=\frac{m_3-m_{ad}}{m_{ad}-m_1}
\end{equation}
Since $m_3=m_1$, $S_3=S_1$ for this simple model.

\section{The Effect of Varying The Thickness of the Stable
Layer}\label{thick}

The primary focus of this paper is to explore the effect of
increasing the width of the convectively stable region, via
increasing the total domain depth $\mathcal{D}$, on strength of the
interaction between the unstable layers. Increasing the width of the
convectively stable zone is the most natural way to increase the
number of pressure scale heights across this region. We begin by
considering the case where  all there zones are equal and so
$\mathcal{D}=3.0$. The resolution for this case is 64:64:400. Our
tests have shown, as is frequently the case in convection simulations, that fewer modes are needed in the $x$ and $y$ directions
than grid points needed in the vertical. We employ this resolution throughout, except in the next section where the height of the box
is varied.

\begin{figure}
\begin{center}
\epsfig{file=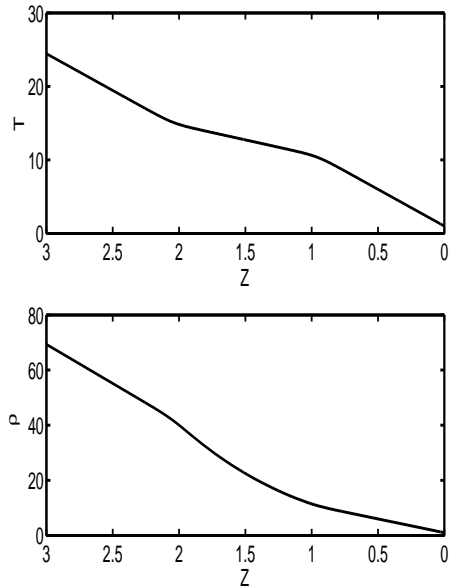,width=6cm,height=8cm}
\end{center}
  \caption{The initial temperature and density profiles when $\mathcal{D}=3.0$.}\label{fidback}
\end{figure}

As we outlined in the previous section, the conductivity is
non-linear in this problem and so we must solve for the
initial density and temperature profile for each case. For the
fiducial case (for which the stable region is the same width as the
unstable regions) the initial temperature and density variations are
shown in Figure \ref{fidback}. As the earlier analytic theory
indicated that the number of pressure scale heights of separation
between the two unstable layers is a crucial factor we will focus on this quantity. For the fiducial case there are 1.60 pressure scale heights across the
convectively stable mid-layer, which is
rather less than the two pressure scale heights estimate suggested
by the earlier analytic theory as necessary for true separation. 

\begin{figure}
\begin{center}
\epsfig{file=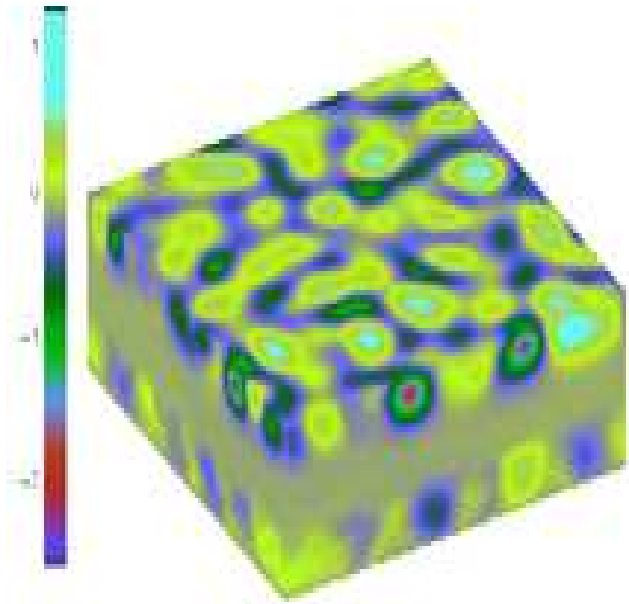,width=10cm,height=10cm}
\end{center}
  \caption{Plot at $t=2.81$
   Sides of the box show the vertical momentum
  flux and the top shows the vertical momentum flux near the top of
  the box for the case where $\mathcal{D}=3.0$ and $S_2=5.0$.}\label{fidearly}
\end{figure}

\begin{figure}
\begin{center}
\epsfig{file=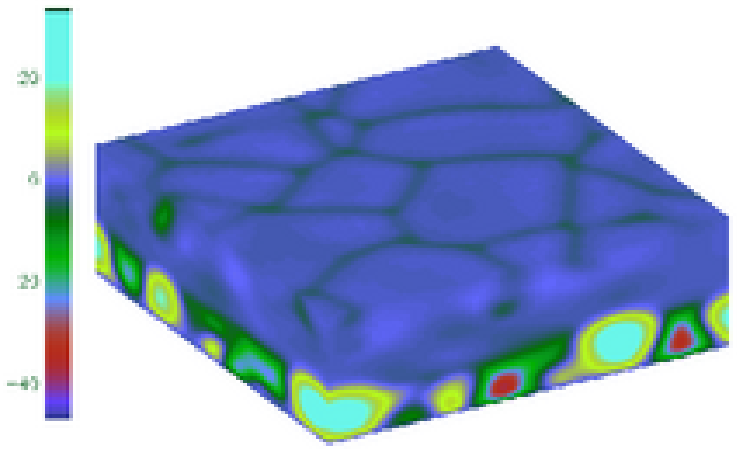,width=10cm,height=10cm}
\end{center}
  \caption{Plot at $t=8.41$ Sides of the box show the vertical momentum
  flux and the top of the box shows the vertical momentum flux near
  the top of the box for the case where $\mathcal{D}=3.0$ and $S_2=5.0$.}\label{fidmid}

\end{figure}

\begin{figure}
\begin{center}
\epsfig{file=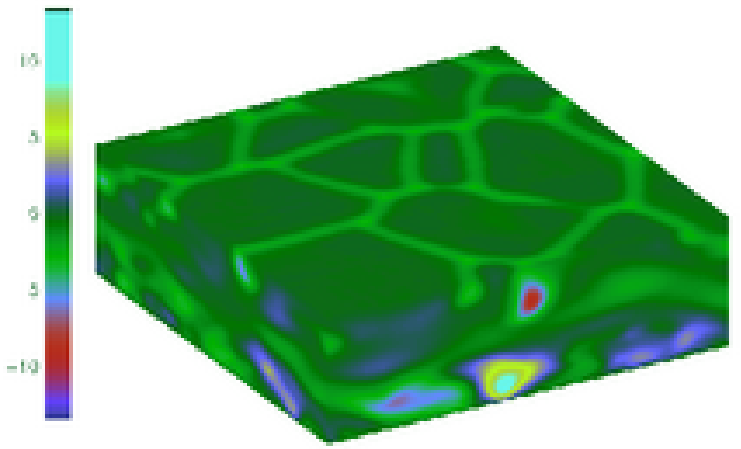,width=10cm,height=10cm}
\end{center}
  \caption{Plot at $t=15.02$ Sides of the box show the vertical momentum
  flux and the top of the box shows the vertical momentum flux near
  the top of the box for the case where $\mathcal{D}=3.0$ and $S_2=5.0$.}\label{fidevolved}
\end{figure}

\begin{figure}
\begin{center}
\epsfig{file=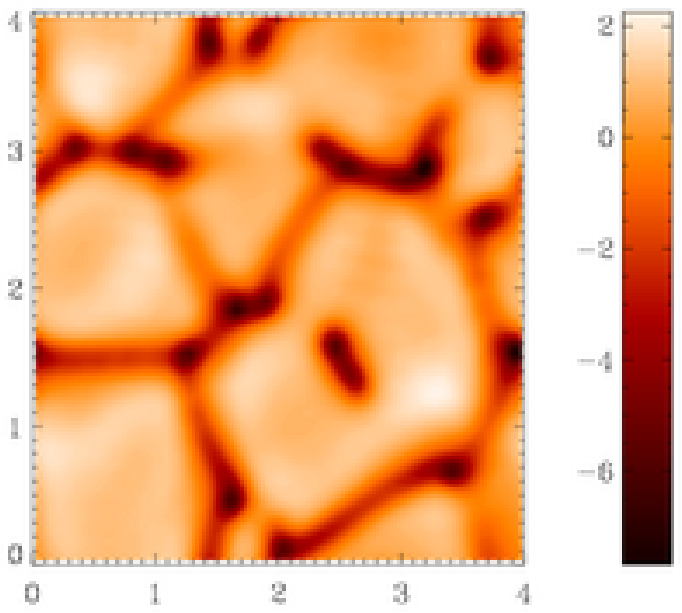,width=8cm,height=6cm}
\epsfig{file=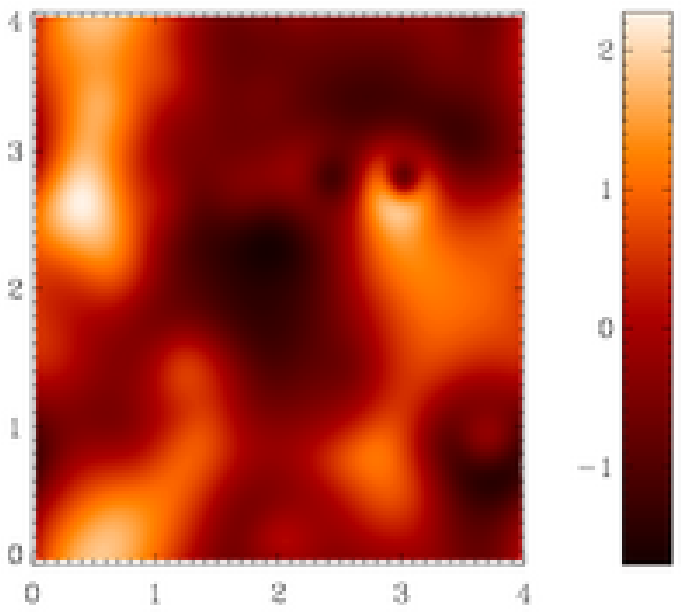,width=8cm,height=6cm}
\epsfig{file=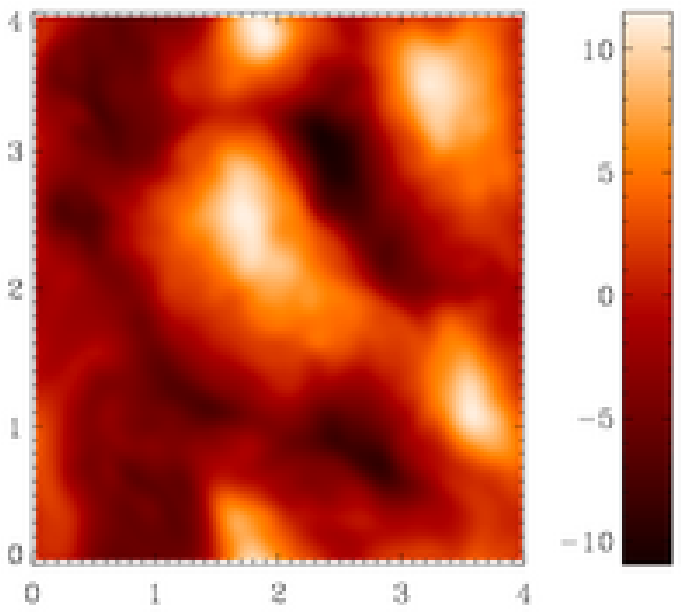,width=8cm,height=6cm}
\end{center}
  \caption{Slices showing the vertical component of momentum through each of the three zones.}\label{slices}
\end{figure}

The state shown in Figure \ref{fidback} is perturbed and allowed to
evolve. At early times, as Figure \ref{fidearly} shows, the motion
is strongest in the upper convectively unstable zone while
comparatively small at the bottom of the box. However, Figure
\ref{fidmid} shows that as time progresses motion in the
lower layer becomes more vigorous.  A
statistically steady state is fully established within a few turnover times although there are significant
temporal fluctuations. This
state is illustrated in Figure \ref{fidevolved} and Figure
\ref{slices} shows horizontal slices though each of the three
zones. As one might expect the convection is
noticeably different in the two convection zones and we find that the average kinetic energy in the middle zone is comparable with that for the
top region but that the energy in the bottom region is three times larger.

While Figure \ref{fidevolved} provides a useful picture of the
convective state it is difficult to gauge
the motions within the convectively stable region from such a
figure. To obtain a clearer picture of the potential interactions
between the two convectively unstable regions it is helpful to note
that, if the two layers are to be considered as `independent' from
each other, then there needs to be a region between the layers where the
velocity becomes very small. Therefore, to facilitate a clearer picture of
the degree of interaction of the layers we calculate the variation in $z$ of the horizontal averages of the modulus of
the $z$ component of momentum.

\begin{figure}
\begin{center}
\epsfig{file=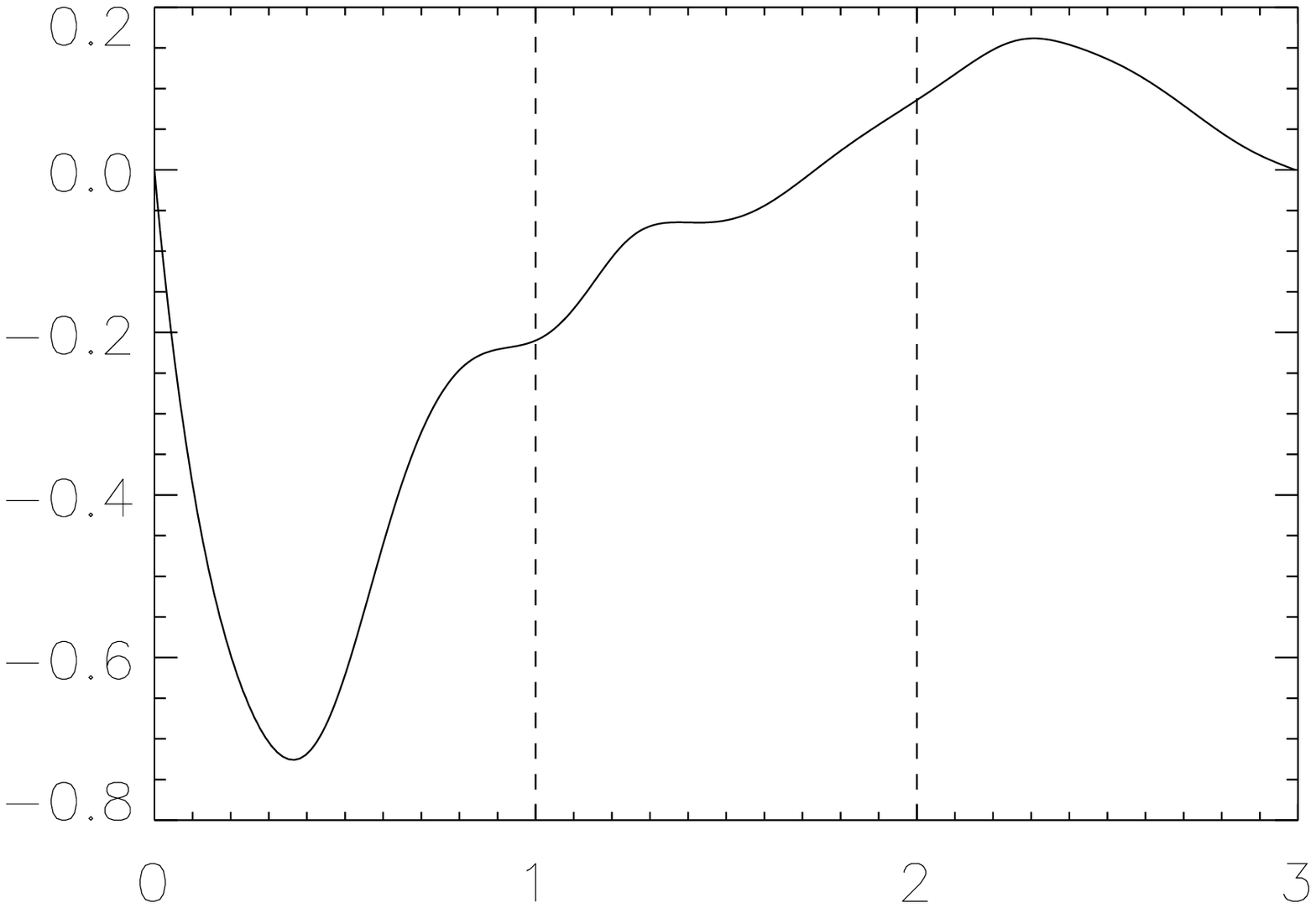,width=8cm,height=8cm}
\end{center}
  \caption{Plot at $t=15.02$. The variation of the vertical component of velocity as a function of height at the centre of the box for the case where $\mathcal{D}=3.0$ and $S_2=5.0$.}\label{vv}
\end{figure}

\begin{figure}
\begin{center}
\epsfig{file=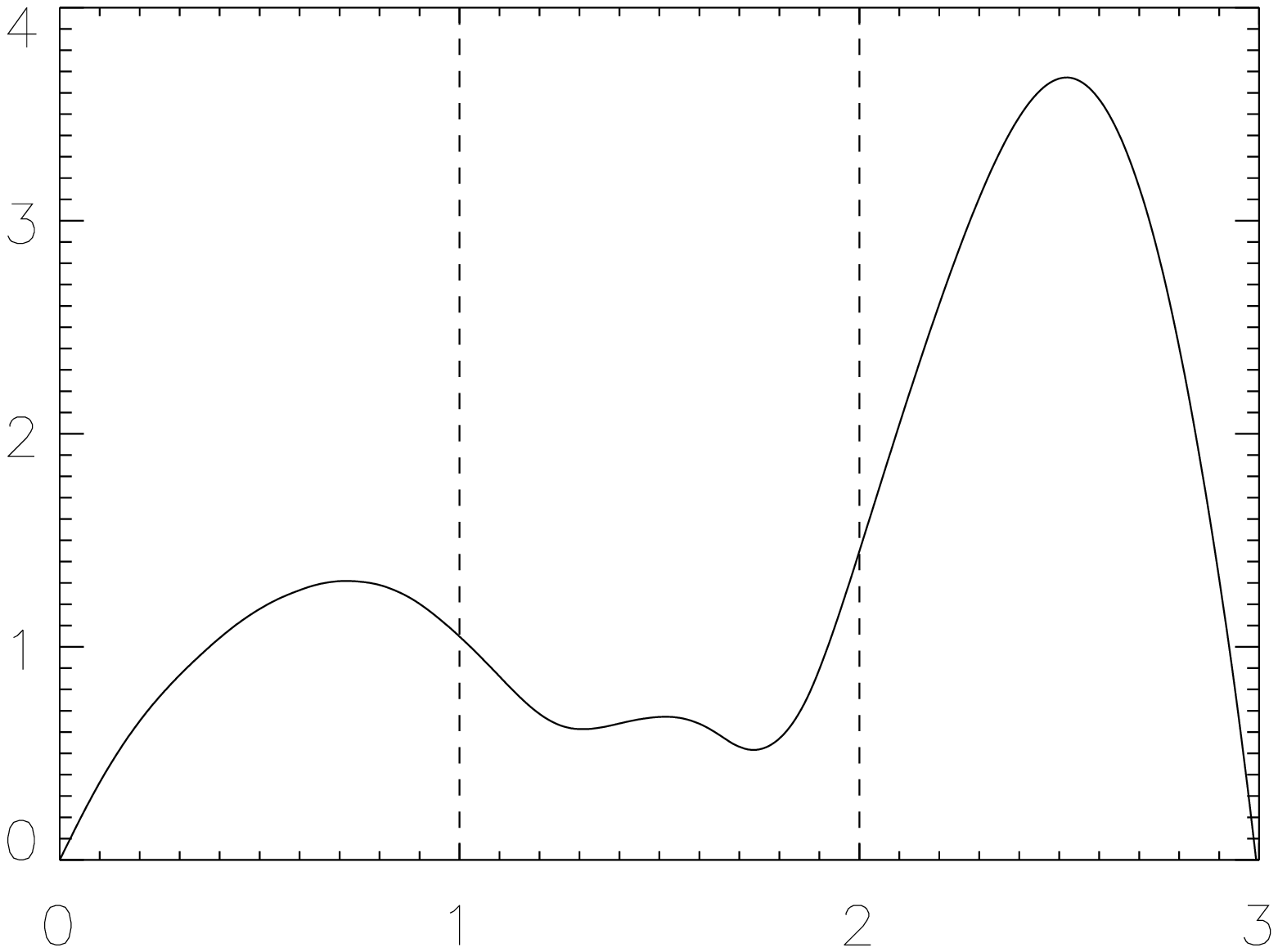,width=8cm,height=8cm}
\end{center} \caption{Variation of
the horizontal average of the modulus of the z-component of momentum
at $t=15.02$ for the case where $\mathcal{D}=3.0$ and $S_2=5.0$.}
\label{fid_zvar}
\end{figure}

Figure \ref{fid_zvar} shows that while there is clearly more vertical motion in the
convectively unstable regions, as one might
anticipate,  there is still a non-negligible
vertical component of momentum in the middle of the box. Thus the two
convection zones are connected in this case and so there  is a conduit for mixing between the two convection
zones for this level of separation. Any reduction in the width of
the convectively stable region will decrease the number of pressure
scale heights of variation across the layer and increase the level
of interaction between the two unstable zones. Such vigorous motion in the convectively stable
layer was discussed in the context of the downward directed hexagon case of Latour \textit{et al.} and is further confirmed
by a plot of the vertical component of velocity as shown in
\ref{vv}. 

Although the Rayleigh number of the convection is very large the numerical resolution is not available to conduct simulations with very large Reynolds numbers; here the peak values are of order 100. 
Improving the resolution, and so using even larger Rayleigh numbers, would serve to increase the amount of interaction of the unstable layers. 

The most obvious way to limit the interaction of the unstable
regions is to increase the layer depth, $\mathcal{D}$, and so increase the
width of the stable region if the convectively unstable regions have
the same height. In this paper we discuss three further  values of
$\mathcal{D}$ namely, 3.5, 4.0 and 5.0, for which the corresponding
pressure scale height variations across the mid-layer are 2.23, 2.80
and 3.77. It is important to note that all three of these cases are
above the two pressure scale heights estimate that suggsted by the
analytic theory as the transition between connected and unconnected convection layers.

\begin{figure}
\begin{center}
\epsfig{file=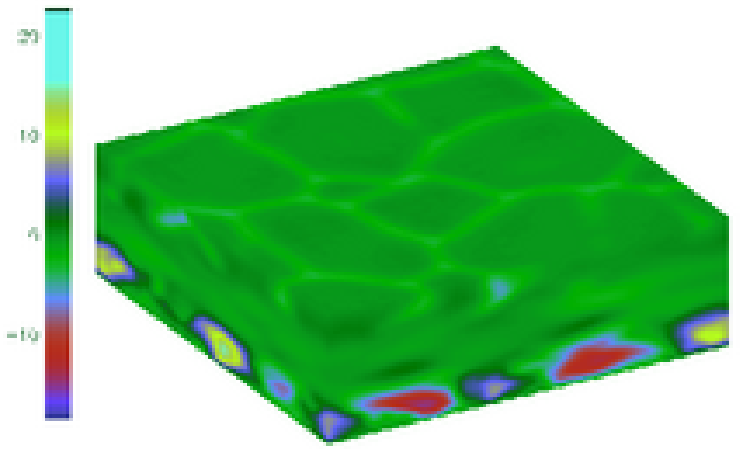,width=10cm,height=10cm}
\end{center}
  \caption{Plot at $t=15.34$ Sides of the box show the vertical momentum
  flux and the top of the box shows the vertical momentum flux near
  the top of the box for the case where $\mathcal{D}=3.5$.}\label{D3}
\end{figure}

Figure \ref{D3} shows a snapshot of the vertical component of the
momentum for the case where $\mathcal{D}=3.5$. This figure is
qualitatively the same as that for $\mathcal{D}=3.0$ and we note
here that similar plots are obtained at larger box heights.

\begin{figure}
\begin{center}
\epsfig{file=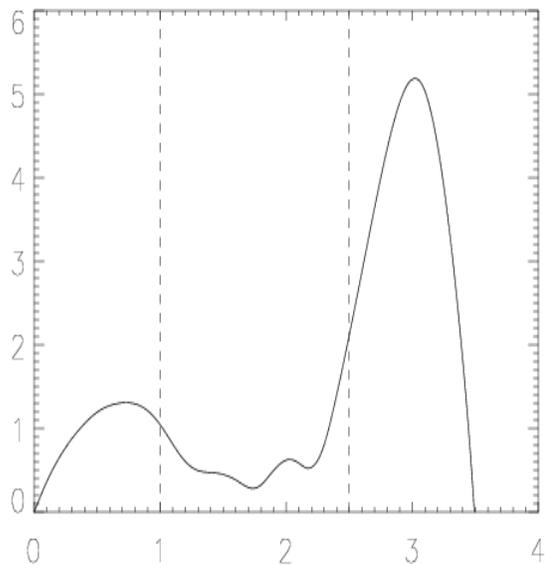,width=8cm,height=8cm}
\end{center}
\caption{Variation of the horizontal average of the modulus of the
z-component of momentum at $t=15.34$ when $\mathcal{D}=3.5$.}
\label{D3a}
\end{figure}

\begin{figure}
\begin{center}
\epsfig{file=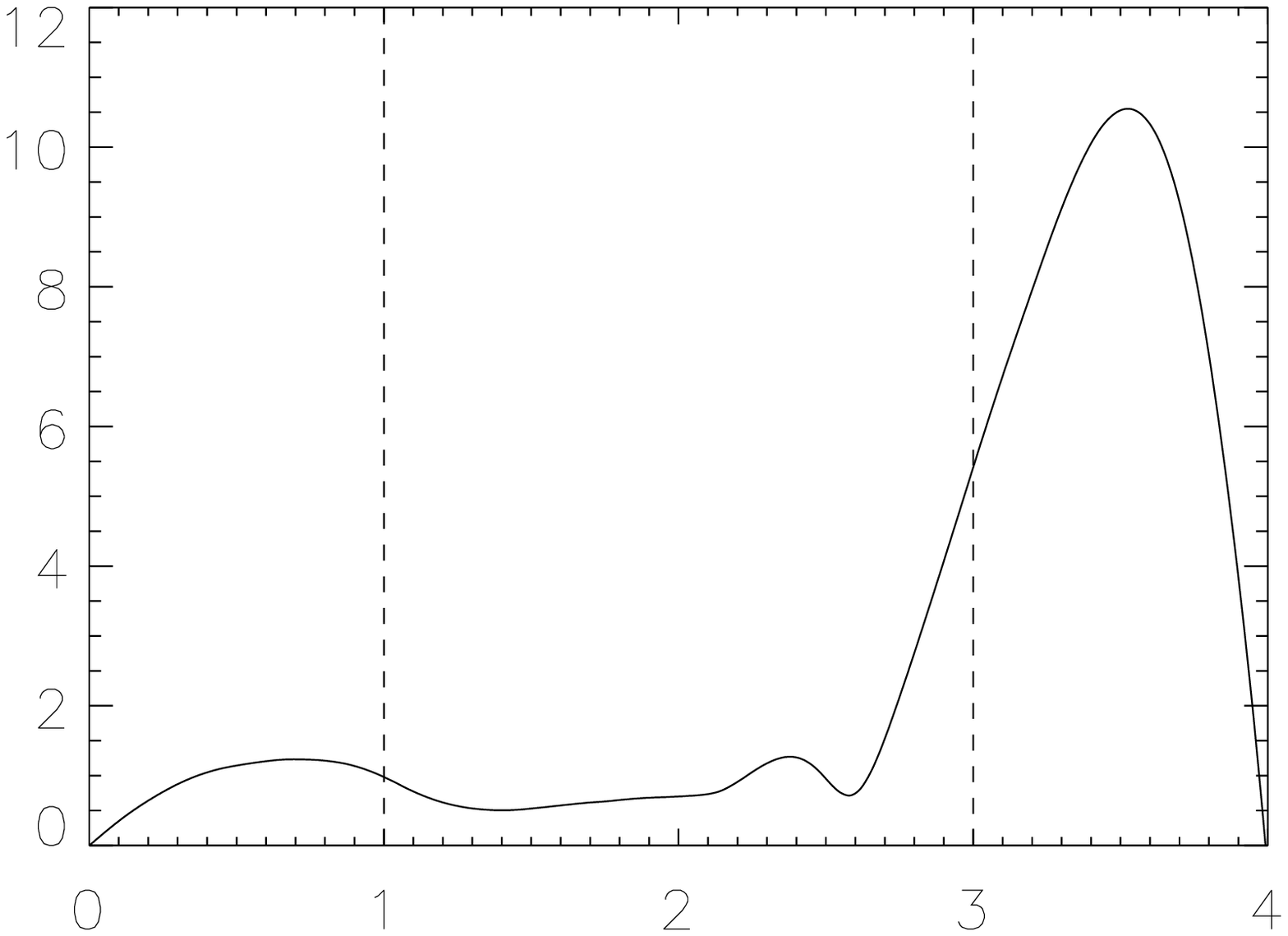,width=8cm,height=8cm}
\end{center}
\caption{Variation of the horizontal average of the modulus of the
z-component of momentum at $t=15.17$ when
$\mathcal{D}=4.0$.}\label{D4a}

\end{figure}

\begin{figure}
\begin{center}
\epsfig{file=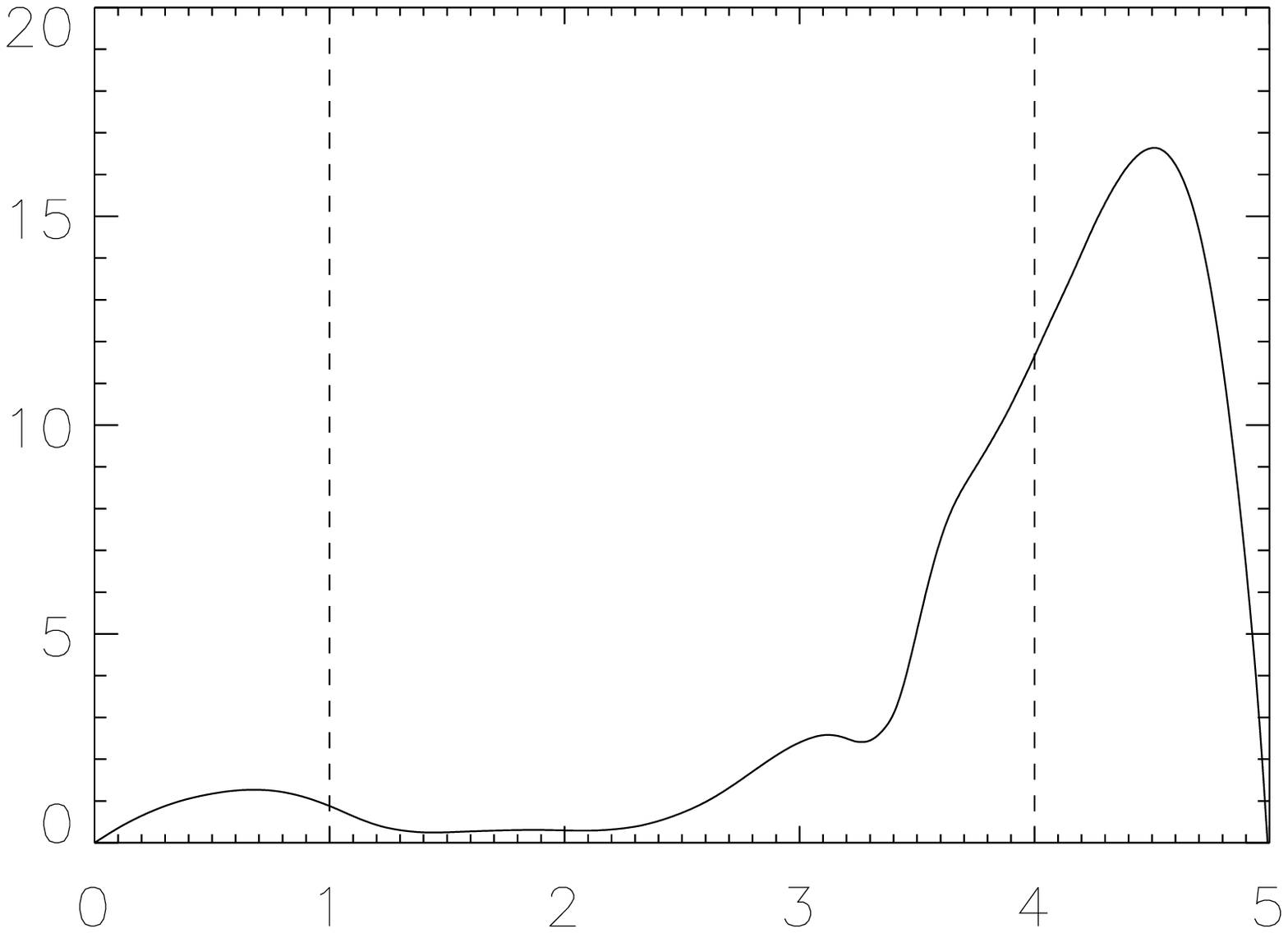,width=8cm,height=8cm}
\end{center}
\caption{Variation of the horizontal average of the modulus of the
z-component of momentum at $t=15.16$ when
$\mathcal{D}=5.0$.}\label{D5a}

\end{figure}

As for the case when $\mathcal{D}=3.0$, we consider the variation of
the planar average vertical component of momentum throughout the
domain in the established statistically steady state, for each of
the increased box heights. Figures \ref{D3a}-\ref{D5a} show this
quantity for each case. Note that there is a change in the magnitude
of the vertical momentum flux in  the bottom zone as the box
height is increased. This is because the density in this lower
region increases as the box height is elongated. This is a follow on
effect of increasing the stable layer. One should note that in the
static state the density and temperature
differences across the upper convection zone remains unchanged. The
increase in the depth of the convectively stable region alters the
difference in these quantities across the stable layer, which implies 
a greater density at the top of the lower convection zone.

Each of these figures clearly illustrates that, while there is a
significant reduction in vertical motion in the convectively stable
region, there is a non negligible vertical component of momentum
throughout the box for all box heights. The motion within the
convectively stable region in all these cases is generated by overshooting plumes from the convectively unstable
regions that lie on either side of this layer. As there is no point
at which the vertical component of momentum vanishes in any of these
cases there is a route for mixing of passive and dynamic quantities
between the two convectively unstable regions. Only for a substantial
increase in the box height will the
vertical component of momentum fall to a very small value at one plane in the box.
However, further increase in the box height would be extremely
computationally expensive for this problem and also unwarranted in
the physical context. In A-type main sequence stars the separation
between the unstable layers does not extend past a couple of scale
heights. Therefore, on the basis of the present work we can conclude
that in A-type stars there is a clear connection between the
convectively unstable zones that lie immediately below the stellar
photosphere.

\section{The Effect of Varying The Stiffness of the Stable Layer}

In the context of A-type stars, the conductivity in the convectively
stable layer is not expected to vary greatly from that in the unstable
zones and so, in the language of the model, the stiffness is low.
However, from a mathematical view point it is interesting to give
some consideration to what happens if the stiffness of the
mid-region is increased. Typical values of the stiffness in the
convectively stable layer in previous numerical simulations have
reached 15 (\cite{TBCT1}; \cite{TBC2}), a very large value. However, here we choose to push up even further to a $S_2$ value of
30; this is certainly greater than that which is encountered in
A-type stars. Even such a large stiffness parameter only yields a variation of 1.91 pressure scale heights
across the stable region. Therefore, the analytic arguments by
Latour \textit{et al.} lead us to expect that, even for such an
extreme choice of the stiffness parameter, there will be some
connection between the two convectively unstable layers.

To show that this conjecture is valid, we  fix the height of the box to 3.0 units and consider three further values of $S_2$ namely, 10
15, and 30. Such values of the mid-layer stiffness give rise to a difference across the mid-layer of 1.72, 1.78 and 1.91
pressure scale heights.

\begin{figure}
\begin{center}
\epsfig{file=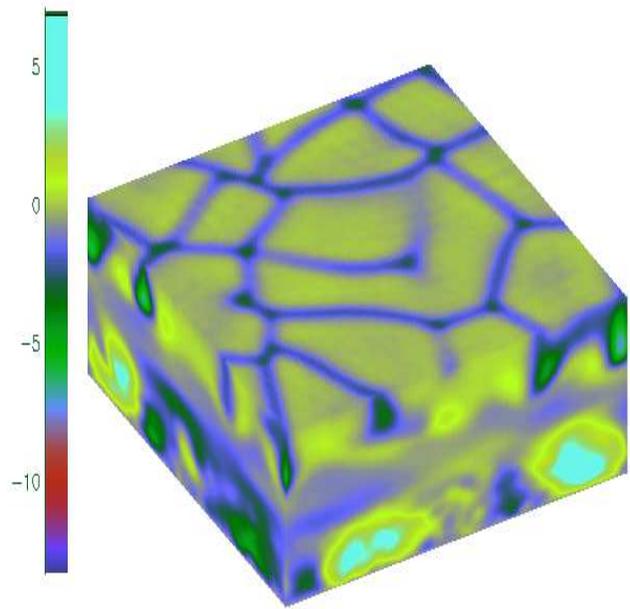,width=10cm,height=10cm}\end{center}
  \caption{Plot at $t=15.12$ Sides of the box show the vertical momentum
  flux and the top of the box shows the vertical momentum flux near
  the top of the box for the case where $\mathcal{D}=3.0$ and $S_2=10$.}\label{K1}
\end{figure}

\begin{figure}
\begin{center}
\epsfig{file=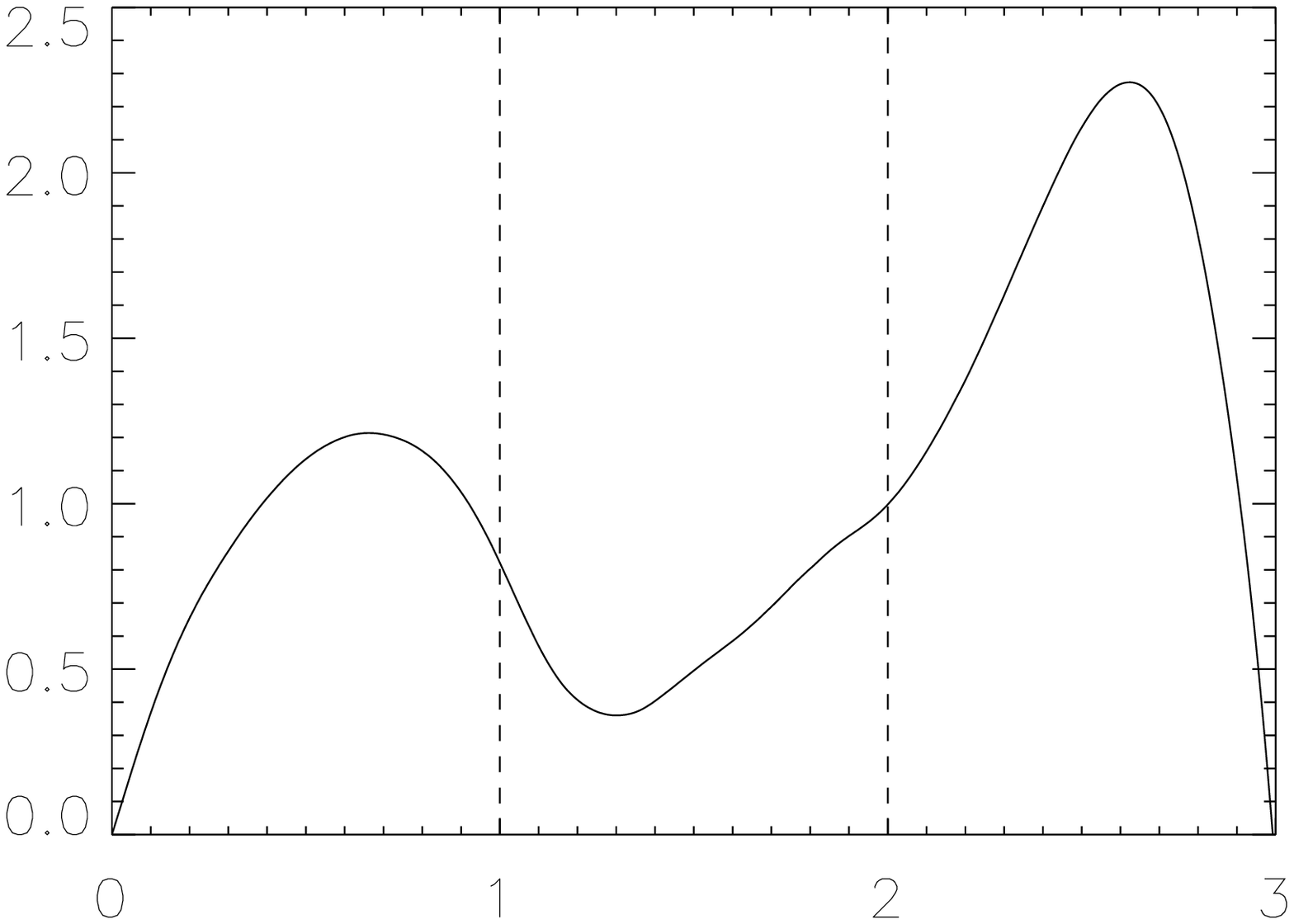,width=8cm,height=8cm}\end{center}
\caption{Variation of the horizontal average of the modulus of the
z-component of momentum at $t=15.12$ in the case where
$\mathcal{D}=3.0$ and $S=10$.}\label{K2}
\end{figure}

\begin{figure}
\begin{center}
\epsfig{file=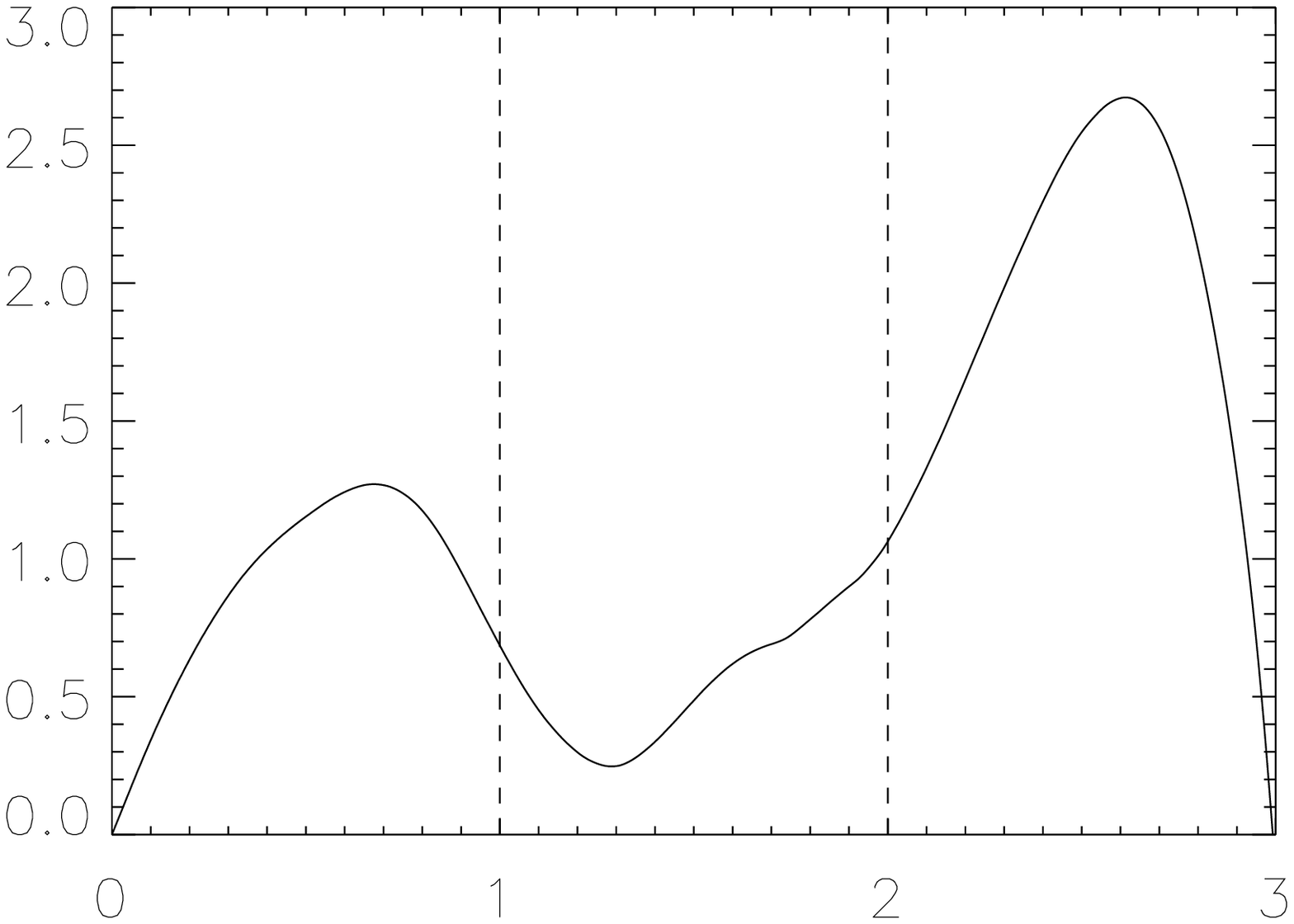,width=8cm,height=8cm}\end{center}
\caption{Variation of the horizontal average of the modulus of the
z-component of momentum at $t=15.17$ for the case where $S=15$.}
\label {K3}
\end{figure}

\begin{figure}
\begin{center}
\epsfig{file=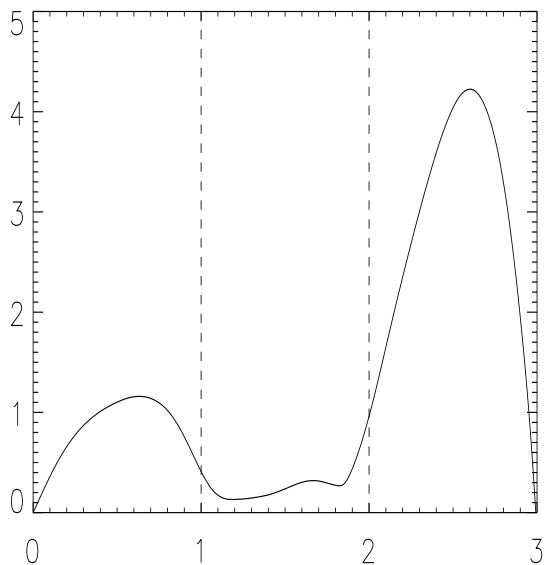,width=8cm,height=8cm}\end{center}
\caption{Variation of the horizontal average of the modulus of the
z-component of momentum at $t=15.16$ for the case where $S=30$.}
\label {K4}
\end{figure}

Figures \ref{K1}-\ref{K4} clearly show that, even for cases with an
extreme mid-layer stiffness, the convectively unstable regions can
not satisfactorily be considered as separate entities, which once
again has important implications for future models that aim to
examine mixing and transport in stars with multiple convection
zones.

\section{Conclusions}

There is still a variety of questions and issues that need to be
fully resolved about the complex dynamical interactions that occur
within A-type stars. The mere presence of multiple convection zones
within these stars implies an extra degree of complexity compared to
that found in G-type stars such as the Sun. In the present paper we focus
on a basic question concerning the conditions that would be required
for the convection zones in A-type stars to be considered as
non-interacting. To this end we considered an idealized model to shed
greater light on the outer convection zones that exist below the
stellar photosphere.

In this paper we focused on two potential ways by which the
interaction between two convection zones could be reduced, namely
the effect of increasing the distance between the two convection
zones and increasing the stiffness on the mid-layer. Both of these
approaches naturally give rise to an increase in the number of
pressure scale heights of variation across the convectively stable
layer that separates the two convectively stable zones, which was
suggested to be an important factor in earlier analytic work.

In section \ref{thick} we examined the effect of increasing the
height of the box from a fiducial case where the box height is such
that both convectively unstable and stable regions have the same
height. As we increased the height of the box we fixed the height of
the unstable layers and therefore a box height increase implies and
increase in the height of the convectively stable layer. Such an
increase naturally drives up the pressure scale height variation
across the stable zone in the middle of the box and so, with this
approach, we are also able to test the earlier theory of Latour
\textit{et al.} that at least two pressure scale heights
of separation is required for the convectively unstable layers not to interact. We showed that even for a box height of five units and
a corresponding pressure variation of almost four scale
heights we did not find that the convectively unstable regions can
be considered as separate, which is not what was suggested by the
earlier analytic theory, though quantitative comparison cannot be
expected from such idealized models. For main-sequence A-type stars it is
unlikely that  there are more than four
pressure scale heights of difference spanning the convectively
stable region that separates the outer two convection zones.
Therefore, this work shows that these two regions will interact and
the interaction will give rise to drastically different mixing and
transport than if they could be considered as separate. It is thus clear that the
convection work that was motivated by a desire to understand
transport in the solar convection zone does not naturally extend to
these stars.

In the second results section we attempted to separate the two
convection zones by increasing the stiffness of the stable layer.
However, we greatly exceeded the level of stiffness that can be
anticipated in the region in A-type stars and we still found that
the two convectively stable regions are connected. This adds further
weight to the point made above that when contemplating the mixing
and transport in A-type stars the convection layers cannot be treated as isolated.

One of the major objectives of this paper is to provide a solid
hydrodynamical basis on which more complex models can be constructed to 
understand fully the dynamics that occur below the surface of A-type
stars. With such a simple model we are not yet in a position to
address important secondary questions such as the influence of the observed chemical anomalies (see discussions in
\cite{Michaud1970} and
\cite{Vauclair1982}  for
more details). However, our model provides a platform on which we can build, so as to address
the effects of rotation, magnetic fields and other
issues related to the dynamics in these stars. We acknowledge here that it has been suggested by some models that the second convection
zone (see, for example, \cite{Vauclair1974} and \cite{Richer1992}) could
vanish under certain conditions although there is clearly no concensus
in the literature. We anticipate that the model could be extended so as to clarify this issue also. 
Work on these extensions is currently in progress.

\section*{Acknowledgments}

LJS wishes to thank the Department of Applied Mathematics and
Theoretical Physics at the University of Cambridge for the award of
a Crighton fellowship for partial support of this research. She would
also like to acknowledge the financial assistance she received via
the Chaire d'Excellence award to Professor Steve Balbus at the Ecole
Normale Sup\'erieure in Paris. We thank Steve Houghton
for his earlier contributions to the numerical code, Douglas Gough for his helpful
comments in the early stages of this work ,and Paul Bushby for many
useful discussions. Finally, we wish to thank the
referee for helpful and constructive comments.

\end{document}